\documentstyle[manuscript,pra,aps]{revtex}
\begin{document}
\title{Free harmonic oscillators, Jack polynomials and Calogero-Sutherland
systems}  
\author{N. Gurappa\thanks{panisprs@uohyd.ernet.in} and Prasanta K.
Panigrahi\thanks{panisp@uohyd.ernet.in}}  
\address{School of Physics, 
University of Hyderabad,
Hyderabad,\\ Andhra Pradesh,
500 046 INDIA.}
\maketitle

\begin{abstract} 
The algebraic structure and the relationships between the
eigenspaces of the Calogero-Sutherland model (CSM) and the Sutherland
model (SM) on a circle are investigated through the Cherednik
operators. We find an exact connection between the simultaneous
non-symmetric eigenfunctions of the $A_{N-1}$ Cherednik operators, from
which the eigenfunctions of the CSM and SM are constructed, and the
monomials. This construction, not only, allows one to write down a
harmonic oscillator algebra involving the Cherednik operators, which
yields the raising and lowering operators for both of these models, but
also shows the connection of the CSM with free oscillators and the SM
with free particles on a circle. We also point out the subtle
differences between the excitations of the CSM and the SM. 

\end{abstract}
\draft

\newpage

\section{Introduction}

Exactly solvable and quantum integrable many-body Hamiltonians with
non-trivial potentials are of great interest, since they lead to a
deeper understanding of the role of interaction in these models. Among
such systems, the Calogero-Sutherland model (CSM) \cite{cal,sut0} and
its generalizations \cite{sut,amp} enjoy a special status and have been
found to be of relevance in diverse branches of physics \cite{al}, such
as the universal conductance fluctuations in mesoscopic systems
\cite{ms1,ms2,ms3}, quantum Hall effect \cite{qhe1,qhe2,qhe3}, 
fluid dynamics \cite{wp}, random matrix theory
\cite{sut0,al,naka,rmt}, fractional statistics \cite{fs1,fs2,fs3},
anyons \cite{anyon1,anyon2,anyon3}, gravity \cite{2dg1,2dg2,2dg3} and
gauge theories \cite{gt1,gt2}. They describe $N$-identical particles in
one-dimension with pair-wise inverse distance-square long-range
interactions. The CSM is defined on the entire real line with or without
the harmonic confinement, while the Sutherland model (SM)
\cite{sut} lives on a circle. The underlying algebraic structure and the
commuting conserved operators of these models have been analyzed by the
exchange operator formalism (EOF) \cite{exop,eop} and the quantum Lax
formulation (QLF)\cite{kh1,kh3,kh4,kh5,kh6}. The EOF makes use of
the well-known Dunkl operator \cite{do}. Both of these models
share the same algebraic structure
\cite{eop,kh1,kh3,kh4,kh5,kh6}. The celebrated Jack polynomials
\cite{jack1,jack2,jack3,jack4} arise as the polynomial part of the
orthogonal basis for the SM and their one parameter deformation, known as
the Hi-Jack polynomials, play the same role for the CSM.
Rodrigues-type formulae have been discovered for the Jack \cite{roj}
and the Hi-Jack polynomials \cite{rohj}. The SM is better understood as
compared to the CSM, since, its exact density-density dynamical
correlation functions have been computed \cite{al,hz,for,les,ha}. The
eigenspectrum of the SM can be interpreted as arising from a set of
free quasi-particles satisfying the generalized exclusion principle
\cite{hald,poly}.  

Yet another approach, towards the understanding of these correlated
systems, is to map them to the corresponding interaction-free systems.
Motivated by Calogero's conjecture \cite{cal1}, the present authors have
shown that the CSM is equivalent to a set of free harmonic oscillators
\cite{prb}. Later, the same model without the harmonic confinement was
found to be unitarily equivalent to free particles on the real line
\cite{gon}; the corresponding eigenfunctions can also be realized as
coherent states \cite{coherent}. Recently, the present authors have
developed a general method, by which, a host of non-trivial interacting
models can be mapped to non-interacting systems \cite{smfpc}. This method
treats both the CSM and the SM on equal footing; the former one is
equivalent to decoupled harmonic oscillators and the later, to free
particles on a circle.  

This paper is organized as follows. The following section deals with the
algebraic structure and the relationships between the Hilbert spaces of
the CSM and the SM, via the Cherednik operators \cite{che}. The origin
of two orthogonal basis sets of the CSM, with two different
innerproducts, is clearly pointed out. In the subsequent section, we
find the exact connection between the eigenfunctions of the Cherednik
operators and the non-symmetric monomials. From this, the raising and
lowering operators for the eigenfunctions of the CSM and the SM are
constructed. Subtle properties of these operators, with correlations
built into them, are pointed out and contrasted with the free
oscillators raising and lowering operators. We conclude in the final
section by pointing out the advantages of the present approach as
compared to the others in the literature and the future directions of
work. The appendix deals with the explicit derivation of the map, which
is used to obtain the connection between the eigenfunctions of the
Cherednik operators and the non-symmetric monomials. 

\section{Algebraic structure of the CSM \& SM, and the relationship
between their Hilbert spaces}

In the following, we first touch upon some common aspects of both the
CSM and the SM, since these are important for concluding the major
results of the present paper. In particular, we show the equivalence of
their underlying algebraic structure through the Cherednik operators;
this naturally brings out the relationships of their respective Hilbert
spaces. 

As mentioned earlier, the EOF and the QLF establish that, both the CSM and
the SM share the same algebraic structure which becomes exactly the same
in the limit $\omega \rightarrow \infty$
\cite{eop,kh1,kh3,kh4,kh5,kh6}. Keeping this in  mind and also
the method developed to map the SM to free particles \cite{smfpc}, we
begin with the $N$-particle CSM Hamiltonian, given by ($\hbar = m = 1$),
\begin{equation} \label{csm}
H_{CSM} = - \frac{1}{2} \sum_{i=1}^N \partial_i^2
+ \frac {1}{2} \omega^2 \sum_{i=1}^N x_i^2 + \frac {1}{2} \beta (\beta
- 1) \sum_{{i,j=1}\atop {i\ne j}}^N \frac {1}{(x_i - x_j)^2}\qquad,
\end{equation}
where, $\partial_i \equiv \partial/\partial x_i $, $\omega$ is the
frequency of the oscillators and $ \beta \ge \frac{1}{2}$ is the coupling
parameter.  

The correlated ground-state of the $H_{CSM}$ is known to be of the form
$\psi_0 = Z G$, where $Z \equiv \prod_{i<j}^N [|x_i - x_j|^\beta (x_i
- x_j)^\delta]$ and $G \equiv \exp\{-\frac{1}{2} \omega \sum_i x_i^2\}$;
here, $\delta = 0$ or $1$ corresponds to the quantization of the CSM,
as bosons or fermions, respectively. Without any loss of
generality, we choose $\delta = 0$. 

Performing a division by $\omega$ and a similarity transformation by 
$\hat T \equiv Z G \exp\{- \frac{1}{2 \omega} \hat A\}$, where $\hat A
\equiv[\frac{1}{2} \sum_i \partial_i^2 + \beta
\sum_{i\ne j} \frac{1}{(x_i - x_j)} \partial_i]$,
Eq. (\ref{csm}) can be brought to the following form \cite{prb},
\begin{equation} \label{deos}
\tilde{H}_{CSM} \equiv {\hat T^{-1}} (H_{CSM}/\omega) \hat T = \sum_i
x_i \partial_i + \frac{1}{2} N (N-1) \beta +
\frac{1}{2} N \qquad,
\end{equation}
which is independent of the parameter $\omega$. Using the following
identity,
$$
\sum_{{i,j}\atop i \ne j} \frac{x_i}{(x_i - x_j)} =
\frac{1}{2} N (N-1) \qquad,
$$
Eq. (\ref{deos}) can be rewritten as 
\begin{equation} \label{che}
\tilde{H}_{CSM} = \sum_i \hat{D}_i + \frac{1}{2} N \qquad,
\end{equation}
where, $\hat{D}_i \equiv D_i + \beta (i - 1) - \beta \sum_{j < i} (1 -
K_{ij})$ is the well-known Cherednik operator for the $A_{N - 1}$ root
system \cite{che,roj,bak}; here, $D_i \equiv x_i \nabla_i $, $\nabla_i
\equiv \partial_i + \beta \sum_j^\prime
\frac{1}{(x_i - x_j)} (1 - K_{ij})$ is nothing but the Dunkl derivative
\cite{do} and $K_{ij}$ is the transposition or exchange operator
\cite{exop,eop,exop2,exop3,exop4,exop5}, whose action on an arbitrary
state can be written as   
$$
K_{ij} |x_1, \cdots , x_i, \cdots , x_j, \cdots , x_N > = |x_1, \cdots
, x_j, \cdots , x_i, \cdots , x_N > \qquad.
$$
It is easy to check that, the Cherednik operators are in involution,
{\it i.e.},
\begin{eqnarray} \label{invo}
[\hat{D}_i \,\,,\,\,\hat{D}_j] = 0 = [\tilde{H}_{CSM} \,\,,\,\,\hat{D}_k ]
\qquad,
\end{eqnarray}
for any $i, j, k = 1, 2, 3, \cdots , N$. Henceforth, we follow 
Ref. \cite{roj} for the notations and Ref. \cite{jack3}, for the
definitions of symmetric functions, ordering of partitions, {\it
e.t.c.}, which are not discussed explicitly in the present paper. 

From Eq. (\ref{invo}), it is clear that, by constructing the
simultaneous eigenfunctions, $\chi_\lambda$, of $\hat{D}_i$: 
\begin{eqnarray} \label{chese}
\hat{D}_i \,\,\chi_\lambda = \delta_i^\lambda \,\,\chi_\lambda \qquad,
\end{eqnarray}
and symmetrizing them, one recovers the eigenfunctions, $J_\lambda$, of
the $\tilde{H}_{CSM}$:
$$
J_\lambda = \sum_P \chi_\lambda \quad; \quad(P \quad \mbox{denotes the 
permutations}) \quad.
$$
A generic form of $\chi_\lambda$, for a given partition of $\lambda$,
can be expressed as,
$$
\chi_\lambda = \hat{m}_\lambda + \sum_{\mu < \lambda} u_{\mu \lambda}
\hat{m}_\mu \qquad, 
$$
where, the non-symmetric monomial function $\hat{m}_\lambda = \prod_i
x_i^{\lambda_i} $ and $u_{\mu \lambda}$'s are some coefficients. In Eq.
(\ref{chese}), the eigenvalues $\delta_i^\lambda = \lambda_i + \beta (N
- i)$ and $\lambda = \sum_i \lambda_i$; $\lambda = 0, 1, 2, \cdots,
\infty$ and $\lambda_i$'s are non-negative integers obeying the
dominance ordering \cite{jack3}. Note that, the monomial symmetric
function is given by $m_{\lambda} = \sum_P \hat{m}_\lambda $. The inverse
similarity transformation by $\hat{T}$ on the eigenstates of
$\tilde{H}_{CSM}$ yields the eigenfunctions of the original $H_{CSM}$.
At this moment, it is important to note that, one can construct the following
conserved operator \cite{ber,roj,bak}:  
\begin{eqnarray} \label{ressm}
\tilde{H}_{SM} \equiv \sum_i \left(\hat{D}_i^2 - (N - 1) \beta
\hat{D}_i \right) + \frac{1}{6} N (N - 1) (N - 2) \beta^2 \qquad,
\end{eqnarray}
which, when restricted to act on the symmetric functions of the
variables, $x_i$'s, yields the differential equation for the Jack
polynomials \cite{jack1,jack2,jack3,jack4}, and is nothing but the
Sutherland Hamiltonian \cite{sut}, which is gauged away by the ground-state 
wavefunction and divided by $(2 \pi/L)^2$. Note that, in the Sutherland 
model, $x_j = \exp\{\frac{2 \pi i \theta_j}{L}\}$, where, $\theta_j$ is 
the location of the $j$-th particle on the circle and $L$ is the length 
of the circumference.  

Since, both the $\tilde{H}_{CSM}$ and the $\tilde{H}_{SM}$ can be
expressed in terms of the same Cherednik operators, $\hat{D}_i$, we also 
conclude that, these two models share the same algebraic structure
\cite{eop,kh1,kh3,kh4,kh5,kh6}. Now, the symmetrized simultaneous
eigenfunctions of the Cherednik operators coincide with the Jack
polynomials, because, $[\tilde{H}_{CSM} \,\,,\,\,\tilde{H}_{SM}] = 0$.
Therefore, the polynomial part of the eigenfunctions of the $H_{CSM}$,
{\it i.e.}, the Hi-Jack polynomials, $j_\lambda$, can be expressed as, 
\begin{eqnarray}
j_\lambda(\{x_i \}, \beta, \omega) = e^{- \frac{1}{2 \omega} \hat{A}}
\,\, J_\lambda(\{x_i \}, \beta) \qquad;
\end{eqnarray}
this is Lassalle's famous exponential formula \cite{sogo,bak}. An
alternate derivation of the Lassalle's formula has been earlier given
by Sogo \cite{sogo}.

It is interesting to note from Eq. (\ref{deos}) that, akin to the
Cherednik operators, one can also choose $x_i \partial_i$ as the $N$
commuting operators, whose simultaneous eigenfunctions are 
$\hat{m}_\lambda$ with eigenvalues $\lambda_i$. By symmetrizing the
$\hat{m}_\lambda$'s, one obtains the monomial symmetric functions as the
eigenfunctions of the $\tilde{H}_{CSM}$ with eigenvalues $\sum_i
\lambda_i + \frac{1}{2} N (N - 1) \beta + \frac{1}{2} N $. Hence,
similar to the Hi-Jack polynomials, another set of eigenstates of the
$H_{CSM}$ can be written as   
\begin{eqnarray}
P_\lambda(\{x_i \}, \beta, \omega) = e^{- \frac{1}{2 \omega} \hat{A}}
\,\, m_\lambda(\{x_i \}) \qquad.
\end{eqnarray}
Performing one more similarity transformation by $\exp\{\frac{1}{4 \omega}
\sum_i \partial_i^2 \} \,\, G^{-1}$, $\tilde{H}_{CSM}$ can be mapped 
to a set of $N$ free harmonic oscillators \cite{prb}. Further, by the
inverse transformation, one can define $a_i^+ = \hat{T} x_i
{\hat{T}}^{-1}$ and $a_i^- = \hat{T} \partial_i {\hat{T}}^{-1}$ as the
creation and annihilation operators: $H_i/\omega = a_i^+ a_i^- +
\frac{1}{2} (N - 1) \beta + \frac{1}{2}$, $[a_i^- \,\,,\,\,a_j] =
\delta_{ij}$ and $[H_i/\omega \,\,,\,\,a_j^{\pm}] = \pm \,\, a_{j}^{\pm}
\delta_{ij}$. Now, the $H_{CSM}$ can be written completely in terms of the
decoupled oscillators: 
\begin{eqnarray}
H_{CSM}/\omega = \sum_{i} a_i^+ a_i^- + \frac{1}{2} N (N - 1) \beta +
\frac{1}{2} N \qquad.
\end{eqnarray}
It can be verified by a direct computation that, the Hi-Jack
polynomials, $j_\lambda$, form an orthogonal basis with respect to the
conventional inner product, whereas the $P_\lambda$'s do not \cite{sogo}.
However, it is clear that, $H_{CSM}$ truly becomes a set of
decoupled harmonic oscillators in the $P_\lambda$ basis, but not in
the $j_\lambda$ basis, since, $P_\lambda$'s can be obtained by the
repeated applications of the commuting creation operators, $a_i^+$, on
the ground-state. $P_\lambda$'s can be made orthogonal by postulating a
new inner product $<<\mu | \lambda> = \delta_{\mu \lambda}$, where, $<<
\mu | = <<0| m_\mu(\{a_i^-\})$, $|\lambda > = m_\lambda(\{a_i^+\}) |0>$ and
$a_i^- |0> = 0 = <<0|a_i^+ $, for $i = 1, 2, 3, \cdots, N$ \cite{prb}.
Explicit construction of this new orthogonal basis was achieved in
Ref. \cite{uji1}.  

\section{Free harmonic oscillators and the Jack polynomials}

In the following, using the properties of the Cherednik operators
\cite{che,roj,bak} and the method developed to map the Sutherland model
to free particles \cite{smfpc}, we obtain the Jack
polynomials akin to the $P_\lambda$'s. The Cherednik operators, along
with certain creation and annihilation operators, are found to obey the
free harmonic oscillator algebra. However, as one naturally expects,
these operators, with correlations built into them,  drastically differ
in their properties, when compared 
with the creation and the annihilation operators for the monomial
symmetric functions. 

Rewriting Eq. (\ref{che}) as,
\begin{eqnarray} \label{reche}
\left(x_i \partial_i - \lambda_i + \hat{B}_i \right) \chi_\lambda = 0
\qquad, 
\end{eqnarray}
where, $\hat{B}_i \equiv \beta \sum_j^\prime \frac{x_i}{(x_i -
x_j)} (1 - K_{ij}) - \beta \sum_{j<i} (1 - K_{ij})- \beta (N + 1 - 2
i)$, the solution is found to be (see the appendix for the proof),
\begin{eqnarray}
\chi_\lambda &=& \sum_{n=0}^\infty (- 1)^n \left[\frac{1}{(x_i
\partial_i - \lambda_i)} \hat{B}_i \right]^n \,\, \hat{m}_\lambda
\nonumber\\  
& \equiv & \hat{g}_\lambda \,\, \hat{m}_\lambda \qquad.
\end{eqnarray}
It can be straightforwardly verified that, $\hat{g}_\lambda$ is, indeed,
independent of the particle index $i$, for a given partition of $\lambda$.
Now, the Jack polynomials can be obtained by simply symmetrizing
$\chi_\lambda$: 
\begin{eqnarray} \label{jack}
J_\lambda = \sum_P \chi_\lambda = \sum_P \hat{g}_\lambda
\hat{m}_\lambda \qquad.
\end{eqnarray}
Comparing the above with an earlier derived formula for the Jack polynomials
\cite{smfpc},  
\begin{eqnarray}
J_\lambda(\{x_i \}) = \sum_{n=0}^\infty (- \beta)^n 
\left[\frac{1}{\sum_i [(x_i \partial_i)^2 - \lambda_i^2]} (\sum_{i<j}
\frac{x_i + x_j}{x_i - x_j}(x_i \partial_i - x _j\partial_j) - \sum_i  
(N + 1 - 2 i) \lambda_i )\right]^n \nonumber\\
\qquad \qquad \qquad \qquad \qquad  \times m_\lambda(\{x_i\}) \qquad,
\nonumber 
\end{eqnarray}
we get the following operator identity, satisfied by the monomial
functions:
\begin{eqnarray}
\sum_P && \left(\left[\frac{1}{(x_i \partial_i - \lambda_i)}
[{\sum_i}^\prime \frac{x_i}{(x_i - x_j)} (1 - K_{ij}) - \sum_{j<i} (1 -
K_{ij}) - (N + 1 - 2 i)] \right]^n \,\, \hat{m}_\lambda \right) =
\nonumber\\ && \qquad = \left[\frac{1}{\sum_i [(x_i \partial_i)^2 -
\lambda_i^2]} [\sum_{i<j} \frac{x_i + x_j}{x_i - x_j}(x_i \partial_i -
x _j\partial_j) - \sum_i (N + 1 - 2 i) \lambda_i ]\right]^n  \sum_P
\hat{m}_\lambda \quad. \nonumber 
\end{eqnarray}

From Eq. (\ref{reche}), it can be verified that
\begin{eqnarray} \label{freemap1}
{\hat{g}_\lambda}^{-1} \hat{D}_i \hat{g}_\lambda = x_i \partial_i +
\beta (N - i) \qquad.
\end{eqnarray}
Due to the above result, Eq. (\ref{ressm}) becomes,
\begin{eqnarray} \label{freemap2}
{\hat{g}_\lambda}^{-1} \tilde{H}_{SM} {\hat{g}_\lambda} = \sum_i
\left((x_i \partial_i)^2 + \beta (N + 1 - 2 i) x_i \partial_i \right) \qquad.
\end{eqnarray}
Eqs. (\ref{freemap1}) and (\ref{freemap2}) depict the equivalence of
the CSM to decoupled oscillators and the SM, to free particles on a
circle. This is yet another proof of their equivalences
\cite{prb,smfpc}. 

By the inverse transformation, the creation and annihilation operators
for the Jack polynomials can be defined:
\begin{eqnarray}
b_{i , \lambda}^+ = \hat{g}_\lambda x_i {\hat{g}_\lambda}^{-1} \quad ;
\quad b_{i , \lambda}^- = \hat{g}_\lambda \partial_i
{\hat{g}_\lambda}^{-1} \quad,
\end{eqnarray}
which satisfy,
\begin{eqnarray}
\hat{D}_i &=& b_{i , \lambda}^+  b_{i , \lambda}^- +  \beta (N - i)
\nonumber\\ 
\mbox{and} \qquad \qquad 
[\hat{D}_i \,\,&,&\,\, b_{j , \lambda}^{\pm}] =
\pm b_{j , \lambda}^{\pm} \quad ; \quad [b_{i , \lambda}^{-}
\,\,,\,\,b_{j , \lambda}^{+}] = \delta_{ij} \quad.
\end{eqnarray}
Note that, $b_{j , \lambda}^{\pm}$ crucially depend on a given
partition of $\lambda$; which, in turn, implies that, each Cherednik
operator can be written in terms of an infinite set of decoupled
oscillators. The ground and excited states can be obtained from,
\begin{eqnarray}
b_{i , \lambda}^- |0 >_\lambda &=& 0 \quad \mbox{for} \quad i = 1,2,
\cdots, N \nonumber\\
\mbox{and} \qquad \qquad \prod_i (b_{i , \lambda}^+)^{\mu_i}
|0>_\lambda &=& |\mu >_\lambda \quad \mbox{with} \quad \sum_i \mu_i =
\mu \quad,  
\end{eqnarray}
respectively. However, all the states, $|\mu >_\lambda$, are not
normalizable except those, for which $\mu = \lambda$, {\it i.e.},
$|\lambda >_\lambda$.  

Recollecting the earlier mapping of the SM to free particles
\cite{smfpc}, 
\begin{eqnarray}
\hat{G}_\lambda^{-1} \bar{H}_{SM} \hat{G}_\lambda =
\sum_i (x_i \partial_i)^2 + \beta \sum_i (N + 1 - 2 i) \lambda_i \quad,
\end{eqnarray}
where, 
\begin{eqnarray}
\hat{G}_\lambda \equiv \sum_{n = 0}^{\infty} (- \beta)^n
\left[\frac{1}{\sum_i [(x_i \partial_i)^2 - \lambda_i^2]}
[\sum_{i<j} \frac{x_i + x_j}{x_i - x_j}(x_i \partial_i - x_j \partial_j
) - \sum_i \lambda_i (N + 1 - 2 i)] \right]^n \quad, \nonumber
\end{eqnarray}
and 
\begin{eqnarray}
\bar{H}_{SM} \equiv \sum_i (x_i \partial_i)^2 + \beta \sum_{i<j}
\frac{x_i + x_j}{x_i - x_j} (x_i \partial_i - x_j \partial_j) \quad,
\nonumber 
\end{eqnarray}
one can also define another set of creation and annihilation
operators for the Jack polynomials:
\begin{eqnarray}
c_{i , \lambda}^+ &=& \hat{G}_\lambda x_i {\hat{G}_\lambda}^{-1} \quad ;
\quad c_{i , \lambda}^- = \hat{G}_\lambda \partial_i
{\hat{G}_\lambda}^{-1} \quad, \nonumber\\
\mbox{and} \qquad \qquad \qquad \qquad [c_{i , \lambda}^{-} \,\,& , & \,\,c_{j , \lambda}^{+}] =
\delta_{ij} \quad. 
\end{eqnarray}
Note that, $c_{i , \lambda}^{\pm}$'s also depend on $\lambda$, but,
unlike $b_{i , \lambda}^{\pm}$'s, they are insensitive to the
permutations of the particle coordinates. In order to have the
normalizable eigenstates, one has to symmetrize the states created by
the  repeated application of $c_{i , \lambda}^{+}$'s on the
ground-state which depends on the $\lambda$ and is annihilated by
$c_{i , \lambda}^{-}$'s. This situation is analogous to the case 
encountered earlier, when the CSM is mapped to free harmonic
oscillators \cite{prb,uji1}. Keeping this in mind, a generic state
can be written as,
\begin{eqnarray}
S_{\mu , \lambda} = m_\mu(\{ c_{i,\lambda}^+\}) \phi_{0 ,
\lambda}(\{x_i\}) \qquad,  
\end{eqnarray}
where, $m_\mu$'s are the monomial symmetric functions \cite{jack3},
and $c_{i,\lambda}^- \phi_{0 , \lambda}(\{x_i\}) = 0$. It can be
verified that, unlike the previous situation \cite{prb,uji1}, all 
these symmetrized states, $S_{\mu , \lambda}$, are still not
normalizable, except those, for which $\mu = \lambda$. In this case,
$S_{\lambda , \lambda}$ coincides with the Jack polynomial
$J_{\lambda}$, {\it i.e.}, $S_{\lambda , \lambda} = J_{\lambda}$.

\section{Conclusions}

In conclusion, we have found a mapping between the eigenstates of the
Cherednik operators and the non-symmetric monomials. This not only
allows one to find the common algebraic structure of these two models,
but also enables one to map the CSM to free harmonic oscillators and
the SM to free particles on the circle. Hence, both these models are
treated on equal footing. The earlier known method of generating Jack
polynomials used the creation operators, which were not commuting
ones, and have to be acted in a particular order on the vacuum
\cite{roj}. The present method does not suffer from these difficulties.

The excited states of the Calogero-Sutherland model (CSM) and the
Sutherland model (SM) can be thought to be arising from the excitations
of an infinite set of free harmonic oscillators, labeled by the
partitions of $\lambda$. In other words, from each set of harmonic
oscillators labeled by the partitions of $\lambda$, one can construct
an infinite number of towers such that, each tower contains an infinite
number of excited states bounded from below. However, from each tower
of these excited states, only one state survives as the normalizable
one, which belongs to the Hilbert space. This rich structure needs
further analysis, which is currently under progress and will be
reported elsewhere.

It worth pointing out again that the excited states of the CSM can be
interpreted in two ways due to the presence of two different inner
products \cite{prb,uji1}. In one case, they arise out of the
decoupled oscillators, where as in the other scenario, they originate
from a correlated system. For the SM, only the later interpretation
seems to be valid.

Finally, we would like to remark that, the present procedure can
also be carried out for the root systems, other than the $A_{N-1}$
\cite{amp}. Extension of these analyses to the higher dimensional
models \cite{hd1,hd2,hd3,hd4} may provide new insights; particularly,
in the context of the two-dimensional systems, this may lead to a
better understanding of some intriguing aspects of the anyons
\cite{anyon1,anyon2,anyon3}. Furthermore, this technique may also throw
new light on the structure of the supersymmetric versions of these
models \cite{susy1,susy2}. 

\acknowledgements

The authors acknowledge useful discussions with Prof. V. Srinivasan and 
Prof. S. Chaturvedi. 

\begin{center}
{\bf APPENDIX}
\end{center}

In the following, we connect the solutions of differential equations,
involving the Dunkl derivatives, to the monomials. For that purpose, we
extend the proof given in Ref. \cite{smfpc} for ordinary differential
equations, to the equations involving Dunkl derivatives. First, we
illustrate the procedure for the single variable case, and then extend
it to the multivariable scenario. 

Consider the most general and arbitrary linear differential equation
\cite{smfpc},  
\begin{eqnarray} \label{ie}
\left(F(D) + \hat{P} \right) y(x) = 0 \quad,
\end{eqnarray}
where, $D \equiv x \frac{d}{dx}$ and $F(D) = \sum_{n = - \infty}^{n
= \infty} a_n D^n $, is a diagonal operator. $\hat{P}$ can be an
arbitrary operator, having a well-defined action in the space 
spanned by $x^n$. Here, $a_n$'s are some parameters. The following
ansatz,   
\begin{eqnarray} \label{an}
y(x) &=& C_\lambda \left \{\sum_{m = 0}^{\infty} (-1)^m
\left[\frac{1}{F(D)} \hat{P} \right]^m \right \} x^\lambda \nonumber\\
&\equiv& C_\lambda \hat{G}_\lambda x^\lambda \qquad,
\end{eqnarray}
is a solution of the above equation, provided, $F(D) x^\lambda = 0$ and
the coefficient of $x^\lambda$ in $y(x) - C_\lambda x^\lambda$ is zero
(no summation over $\lambda$); here, $C_\lambda$ is a constant. The
later condition not only guarantees that, the solutions, $y(x)$'s, are
non-singular, but also yields the eigenvalues.

Substituting Eq. (\ref{an}), modulo $C_\lambda$, in Eq. (\ref{ie}),
\begin{eqnarray}
\left(F(D) + \hat{P} \right)&& \left\{\sum_{m = 0}^{\infty}
(-1)^m 
\left[\frac{1}{F(D)} \hat{P} \right]^m \right \} x^\lambda = \nonumber\\
= F(D)&& \left[1 + \frac{1}{F(D)} \hat{P} \right] \left \{\sum_{m =
0}^{\infty} (-1)^m \left[\frac{1}{F(D)} \hat{P} \right]^m \right \}
x^\lambda \nonumber  \\
= F(D)&& \sum_{m = 0}^{\infty} (-1)^m
\left[\frac{1}{F(D)} \hat{P} \right]^m  x^\lambda \nonumber\\
&& + F(D) \sum_{m = 0}^{\infty}(-1)^m \left[\frac{1}{F(D)} \hat{P}
\right ]^{m + 1} x^\lambda \nonumber \\
= F(D)&& x^\lambda - F(D) \sum_{m = 0}^{\infty} (-1)^m
\left[\frac{1}{F(D)} \hat{P} \right]^{m + 1} x^\lambda \nonumber\\
&& + F(D) \sum_{m = 0}^{\infty}(-1)^m \left[\frac{1}{F(D)} \hat{P}
\right ]^{m + 1} x^\lambda \nonumber \\
= 0 \qquad.
\end{eqnarray}
Note that, the detailed properties of $\hat{P}$ are not needed to
prove Eq. (\ref{an}) as a solution of Eq. (\ref{ie}). However,
naturally, they are required while constructing the explicit solutions of
any given linear differential equation.

Eq. (\ref{an}), which connects the solutions of a differential
equation to the monomials, can be generalized to many-variables as
follows \cite{smfpc}.  

Consider,
\begin{eqnarray}
\left(\sum_{n = -\infty}^\infty b_n (\sum_i D_i^n) + \hat{A} \right)
Q_\lambda(\{x_i\}) = B_\lambda(\{x_i\}) \qquad,
\end{eqnarray}
where, $b_n$'s are some parameters, $D_i \equiv x_i \partial_i$,
$\hat{A}$ can be a function of $x_i$, $\partial_i$ and also some other
well-defined operators like the transposition operator $K_{ij}$ and  
$B_\lambda(\{x_i\})$ is a source term. 

\noindent {\it Case (i)} When $B_\lambda(\{x_i\})= 0$ and $\hat{A}
m_\lambda = 
\epsilon_\lambda m_\lambda + \sum_{\mu < \lambda} C_{\mu \lambda}
m_{\mu}$; where, $m_\lambda$'s are the monomial symmetric functions
\cite{jack3} and $\epsilon_\lambda$ and $C_{\lambda \mu}$ are some
coefficients.

Using Eq. (\ref{an}), the solutions can be obtained as,
\begin{eqnarray}
Q_\lambda(\{x_i\}) = \sum_{r=0}^\infty (- 1)^r \left[\frac{1}{((\sum_{n =
-\infty}^\infty b_n (\sum_i D_i^n) - (\sum_{n = -\infty}^\infty b_n
(\sum_i {\lambda}_i^n))} (\hat{A} - \epsilon_\lambda)\right]^r
m_\lambda(\{x_i\}) \,\,
\end{eqnarray}
with, $\sum_{n = -\infty}^\infty b_n (\sum_i {\lambda}_i^n) + 
\epsilon_ \lambda = 0$.

\noindent {\it Case (ii)} When $B_\lambda(\{x_i\}) \ne 0$.

\begin{eqnarray}
Q_\lambda(\{x_i\}) = \sum_{r=0}^\infty (- 1)^r \left[\frac{1}{((\sum_{n
= -\infty}^\infty b_n (\sum_i D_i^n) - (\sum_{n = -\infty}^\infty b_n
(\sum_i {\lambda}_i^n))} \hat{A}\right]^r \nonumber\\
\times \left[\frac{1}{((\sum_{n = -\infty}^\infty b_n (\sum_i D_i^n) -
(\sum_{n = -\infty}^\infty b_n 
(\sum_i {\lambda}_i^n))}\right] B_\lambda(\{x_i\}) \,\,,
\end{eqnarray}
provided, the coefficient of the divergent part in the right hand side
of the above equation is zero. As mentioned earlier, this requirement
yields the eigenvalues.

As an example, consider the hypergeometric differential equation
\cite{gra}, 
\begin{eqnarray}
\left(x (1 - x) \frac{d^2}{d x^2} + [\gamma - (\alpha + \beta + 1) x ]
\frac{d}{d x} - \alpha \beta \right) y(x) = 0 \quad.
\end{eqnarray}
Multiplying by $x$, the above can be written as,
\begin{eqnarray}
\left(D (D + \gamma - 1) - \hat{A}\right) y(x) = 0 \quad,
\end{eqnarray}
where, $D \equiv x \frac{d}{d x}$ and
$\hat{A} \equiv  x (D + \alpha) (D + \beta)$. Now,
$F(D) x^\lambda = D (D + \gamma - 1) x^\lambda = 0$ yields $\lambda
= 0$ or $1 - \gamma$. From Eq. (\ref{an}), the two linearly
independent solutions, $y_0(x)$ and $y_{1 - \gamma}(x)$, can be
written as,  
\begin{eqnarray}
y_0(x) &=& C_0 \left \{\sum_{m = 0}^{\infty} (-1)^m
\left[\frac{1}{D (D + \gamma - 1)} (- \hat{A}) \right]^m \right \}
x^0 \nonumber\\ 
&=& C_0 \,\,e^{\frac{1}{D + \gamma - 1} \hat{A}} \,\,1 \quad,
\end{eqnarray}
and 
\begin{eqnarray}
y_{1 - \gamma}(x) = C_{1 - \gamma}\,\, e^{(1/D) \hat{A}}
\,\,x^{1 - \gamma} \quad.
\end{eqnarray}

Solutions for many other differential equations, constructed by this
method, can be found in Ref. \cite{smfpc}. Further, this technique
can be applied to the bound-state problems of the Schr\"odinger
equations with complicated potentials. These details are also available
in the same reference.

\end{document}